\documentclass[aps,prl,preprint,groupedaddress]{revtex4-1}
\usepackage{amsmath}
\usepackage{amssymb}
\usepackage{graphicx}
\usepackage{epsf}
\usepackage{mathtools}
\begin{document}
\title{Von Neumann entropy in a dispersive cavity}
\author{Ram Narayan Deb}
\email[]{debramnarayan1@gmail.com, debram@rediffmail.com}
\affiliation{Chandernagore College, Chandernagore, Hooghly, PIN-712136, West Bengal, India}
\date{\today}

\begin{abstract}
We study the von Neumann entropy of the partial trace of a system of two two-level atoms (qubits) in a dispersive cavity where the atoms are interacting collectively with a single mode electromagnetic field in the cavity. We make a contrast of this entanglement entropy with the spin squeezing property of the system. We find a close relationship between this von Neumann entropy and the spin squeezing of the system. We observe that this entanglement entropy attains its maximum value when the spin squeezing in the system goes to its maximum and is zero when the spin squeezing in the system vanishes. 
\end{abstract}


\maketitle
\section{I. INTRODUCTION}
Von Neumann entropy, also called the entanglement entropy, forms an important measure of quantum entanglement in bipartite states of quantum systems \cite{Nielsen}.

Optical cavity, in which the atoms interact with the electromagnetic field, is an important experimental tool through which we can study the various aspects of the atom-field interaction and its dynamics. Cavity QED and cavity related systems  are widely studied for producing correlated photons or two-level atoms or both. Dispersive cavity plays a significant role in this field $[2-14]$.

Spin squeezing of a system of two-level atoms is an important experimentally measurable quantity that depends on the amount of quantum correlations established among the atoms \cite{Kitagawa}, 
\cite{Wineland}. So, there exists a close relationship between the von Neumann entropy of the partial trace of a system of two two-level atoms and spin squeezing in that system.
The problem of finding the entropy for the interaction of a single atom and a quantized field in a dispersive cavity has been considered in Ref. \cite{Moya}.

In this paper we consider a system of two identical two-level atoms each of frequency $\omega_0$ interacting collectively with a single mode electromagnetic field in a dispersive cavity whose characteristic frequency nearest to $\omega_0$ is $\omega_c$ \cite{Agarwal}. 
The cavity temperature is related with the average number of thermal photons $\bar{n}$ present in the cavity. The photon lifetime is represented by 
$\frac{1}{2k}$, where $k$ is the cavity bandwidth. 
If $\hat{J}_{+}$, 
$\hat{J}_{-}$ and $\hat{J}_{z}$ are the atomic operators and $\hat{a}$, $\hat{a}^\dagger$ are the bosonic operators of the cavity field mode then the atom-field interaction is governed by the Hamiltonian \cite{Agarwal}
\begin{equation}
\hat{H}_{af} = \hbar\omega_0\hat{J}_z + \hbar\omega_c
\hat{a}^\dagger\hat{a} + \hbar g(\hat{J}_+\hat{a} + \hat{a}^\dagger\hat{J}_{-}),
\label{1a}
\end{equation}
 where $g$ is the atom-field coupling constant. 
                                             
We consider the cavity to be highly detuned, that is, 
$\delta = \omega_c - \omega_0$ is very large satisfying 
$|i\delta + k| >> g\sqrt{2}$.
The solution to the master equation for a quantized cavity mode has been described in Ref. \cite{Arevalo}. Quantum state reconstruction in the presence of dissipation has been described in Ref. \cite{Moya1}, that takes into account cavity losses. Now, for the present case the derivation of the effective Hamiltonian describing the dynamics of the atoms in the cavity, using the master equation approach, has been done by Agarwal et al. in Ref.                                                                          \cite{Agarwal}.  If $\delta >> k$, then the contribution due to damping is very small and the effective Hamiltonian 
 takes the form \cite{Agarwal}
\begin{eqnarray}
\hat{H}_{eff} = \hbar\Delta_{0} [\hat{J}_{+}\hat{J}_{-} + 
2\bar{n}\hat{J}_{z}]
= \hbar\Delta_{0}[\hat{J}^2 - \hat{J}_{}z^2 + 
2\bar{n}\hat{J}_{z}],
\end{eqnarray}
where
\begin{equation}
\Delta_{0} = \frac{g^2\delta}{k^2 + \delta^2}.
\label{1.1}
\end{equation}

This Hamiltonian is quadratic in population inversion operator $\hat{J}_{z}$.  The temperature dependent term in the Hamiltonian produces a simple rotation of the quantum state that has no physical significance. Therefore, we drop it by setting $\bar{n} = 0$. So, the effective Hamiltonian reduces to
\begin{equation}
\hat{H}_{eff} = \hbar\Delta_{0}(\hat{J}^2 - 
\hat{J}_{z}^2).
\label{1.2}
\end{equation}
We consider the initial state of the two two-level atoms, before they enter the cavity, to be an atomic coherent state 
\cite{Radcliffe}, \cite{Arecchi} and after they enter the dispersive cavity the time evolution of the state
is governed by the above Hamiltonian in Eq. (\ref{1.2}). The resulting state is an
atomic Schr$\ddot{o}$dinger cat state \cite{Agarwal}. In this paper we investigate the dynamics of the von Neumann entropy of partial trace of this state and make a contrast with the spin squeezing dynamics of the system.

The organization of the paper is as follows. In Sec. II
 we study the von Neumann entropy of the partial traces of the two two-level atoms in the dispersive cavity. In Sec. III we make a contrast of this entanglement entropy with the spin squeezing dynamics of the whole system.

\section{II. Von Neumann entropy of partial traces of two two-level atoms in a dispersive cavity}

We assume the initial state of the two two-level atoms, before they enter the dispersive cavity,  to be an atomic coherent state which is represented as                     
\begin{equation}
|j,\chi\rangle = \frac{1}{(1 + \chi^2)^{j}}  
\sum_{n=0}^{2j}\sqrt{{}^{2j}C_{n}}\chi^n|j, j-n\rangle.
\label{2.1}
\end{equation}
The quantum number $j = N/2$, where $N$ is the number of atoms.
Since, in this case the number of atoms $N = 2$, the quantum number $j = \frac{N}{2} = 1$. So, the above state reduces to
\begin{equation}
|1,\chi\rangle = \frac{1}{(1 + \chi^2)}\big{[} |1,1\rangle
+ \sqrt{2}\chi|1,0\rangle + \chi^2|1,-1\rangle\big{]}.
\label{2.2}
\end{equation} 

Here $\chi = \tan(\theta/2)e^{i\phi}$. $\chi =0$ represents all the atoms in their upper states. In the following we assume $\phi = 0$ as it has no effect on the problem we study here. This state is achieved by, first, sending the atoms in their upper states through an auxillary cavity. The quantity $\theta$ is decided by the duration of atom-field interaction in this cavity. These atoms then enter the dispersive cavity where their time evolution is governed by the Hamiltonian in Eq. (\ref{1.2}). 

In the $\{m_1, m_2\}$ representation, the time evolved state (with $\hbar = 1$) is
\begin{eqnarray}
|1,\chi,t\rangle &=& \frac{1}{(1 + |\chi|^2)} \bigg{[} 
e^{-i\Delta_0 t}\bigg{|}\frac{1}{2},\frac{1}{2}\bigg{\rangle}
+  
\chi e^{-2i\Delta_0 t}\bigg{|}\frac{1}{2},-\frac{1}{2}\bigg{\rangle} + \chi e^{-2i\Delta_0 t}\bigg{|}-\frac{1}{2},\frac{1}{2}\bigg{\rangle} 
\nonumber\\ 
&+&\chi^2 e^{-i\Delta_0 t}\bigg{|}-\frac{1}{2},-\frac{1}{2}\bigg{\rangle}\bigg{]}.
\label{2.3}
\end{eqnarray}

Now, if $\hat{J}_{1_x}$, $\hat{J}_{1_y}$ and $\hat{J}_{1_z}$ are the operators of atom 1 and 
$\hat{J}_{2_x}$, $\hat{J}_{2_y}$ and 
$\hat{J}_{2_z}$ are the operators of atom 2, then, 
the mean spin vectors of atoms 1 and 2 are
\begin{equation}
\langle\hat{\overrightarrow{J}}_1\rangle = 
\langle\hat{J}_{1_x}\rangle\hat{i} + 
\langle\hat{J}_{1_y}\rangle\hat{j} +
\langle\hat{J}_{1_z}\rangle\hat{k}
\label{2.3a1}
\end{equation} 
and
\begin{equation}
\langle\hat{\overrightarrow{J}}_2\rangle = 
\langle\hat{J}_{2_x}\rangle\hat{i} + 
\langle\hat{J}_{2_y}\rangle\hat{j} +
\langle\hat{J}_{2_z}\rangle\hat{k},
\label{2.3a2} 
\end{equation} 
where $\hat{i}$, $\hat{j}$ and $\hat{k}$ are the unit vectors along the positive $x$, $y$ and $z$ axes respectively.

Now, the density matrix for a pair of qubits $A$ and 
$B$ can be written in the Bloch-Fano decomposition as $[22-25]$
\begin{eqnarray}
\rho_{AB} = \frac{1}{4}\bigg{(} {\bf 1}\otimes
{\bf 1} + \overrightarrow{\sigma}^A.{\bf u}\otimes{\bf 1} + {\bf 1}\otimes\overrightarrow{\sigma}^B.{\bf v} + 
 \sum_{i,j=1}^{3}\beta_{ij}\sigma_i\sigma_j \bigg{)},
\label{1} 
\end{eqnarray}
where
\begin{eqnarray}
{\bf u} &=& \langle\sigma_{x}^{A}\rangle\hat{i} +
\langle\sigma_{y}^{A}\rangle\hat{j} +
\langle\sigma_{z}^{A}\rangle\hat{k},\\
{\bf v} &=& \langle\sigma_{x}^{B}\rangle\hat{i} +
\langle\sigma_{y}^{B}\rangle\hat{j} +
\langle\sigma_{z}^{B}\rangle\hat{k}.
\label{2} 
\end{eqnarray}
From Eq. (\ref{1}), we can obtain two reduced density matrices
\begin{eqnarray}
\rho_A &=& Tr_{B}(\rho_{AB}) = \frac{1}{2}
({\bf 1} + \overrightarrow{\sigma}^A.{\bf u})\\
\rho_B &=& Tr_{A}(\rho_{AB}) = \frac{1}{2}
({\bf 1} + \overrightarrow{\sigma}^B.{\bf v}).
\label{3}
\end{eqnarray}
For pure states $|{\bf u}| = |{\bf v}|$. The well known measure of concurrence, for pure states, can be calculated as
\begin{eqnarray}
C = \sqrt{2[1 - Tr{(\rho^A)}^{2}]} = \sqrt{2[1 - Tr{(\rho^B)}^{2}]}
= \sqrt{1 - |{\bf u}|^2} = \sqrt{1 - |{\bf v}|
^2}.
\end{eqnarray} 
The entanglement of formation is given by
\begin{eqnarray}
F(C) = H\bigg{(}\frac{1 + \sqrt{1 - C^2}}{2}\bigg{)}
 = H\bigg{(}\frac{1 + \sqrt{|{\bf u}|^2}}{2}\bigg{)},
\label{4}
\end{eqnarray}
where
\begin{equation}
H(x) = -x log_2 x - (1 - x)log_2 (1 - x).
\label{5}
\end{equation}
Now, for pure states the entanglement of formation is same as the von Neumann entropy of the reduced density matrix of either $A$ or $B$.
So, using Eqs. (\ref{4}) and (\ref{5}) the von Neumann entropy of the reduced 
density matrix takes the form
\begin{eqnarray}
E = -\bigg{(}\frac{1}{2} + \frac{1}{2}|{\bf u}|\bigg{)} log_2 \bigg{(}\frac{1}{2} + \frac{1}{2}|{\bf u}|\bigg{)}
- \bigg{(}\frac{1}{2} - \frac{1}{2}|{\bf u}|\bigg{)}log_2 \bigg{(}\frac{1}{2} - \frac{1}{2}|{\bf u}|\bigg{)}.
\label{6}
\end{eqnarray}

Now, the mean spin vector for atom 1 is related with $\langle\overrightarrow{\sigma}_1\rangle$ (with $\hbar = 1$) as
\begin{eqnarray}
\langle\hat{\overrightarrow{J_1}}\rangle = 
\langle\hat{J}_{1_x}\rangle\hat{i} +
\langle\hat{J}_{1_y}\rangle\hat{j} +
\langle\hat{J}_{1_z}\rangle\hat{k}
= \frac{1}{2} \bigg{(}\langle{\sigma}_{1_x}\rangle\hat{i} +
\langle{\sigma}_{1_y}\rangle\hat{j} +
\langle{\sigma}_{1_z}\rangle\hat{k}\bigg{)}
= \frac{1}{2}\langle\overrightarrow{\sigma}_1\rangle,  
\end{eqnarray}

So, the von Neumann entropy of the partial traces of the system can be represented in terms of these mean spin vectors of either atom 1 or atom 2 and is given as $[9-12]$
\begin{eqnarray}
E = -\bigg{(}\frac{1}{2} + |\langle\hat{\overrightarrow{J}}_1\rangle| \bigg{)} log_2
\bigg{(}\frac{1}{2} + |\langle\hat{\overrightarrow{J}}_1\rangle| \bigg{)}
- \bigg{(}\frac{1}{2} - |\langle\hat{\overrightarrow{J}}_1\rangle| \bigg{)} log_2
\bigg{(}\frac{1}{2} - |\langle\hat{\overrightarrow{J}}_1\rangle| \bigg{)}
\label{2.3a3} 
\end{eqnarray}
or,
\begin{eqnarray}
E = -\bigg{(}\frac{1}{2} + |\langle\hat{\overrightarrow{J}}_2\rangle| \bigg{)} log_2
\bigg{(}\frac{1}{2} + |\langle\hat{\overrightarrow{J}}_2\rangle| \bigg{)}
- \bigg{(}\frac{1}{2} - |\langle\hat{\overrightarrow{J}}_2\rangle| \bigg{)} log_2
\bigg{(}\frac{1}{2} - |\langle\hat{\overrightarrow{J}}_2\rangle| \bigg{)}.
\label{2.3a4} 
\end{eqnarray}

\begin{figure}
\begin{center}
\includegraphics[width=10.0cm, angle=0]{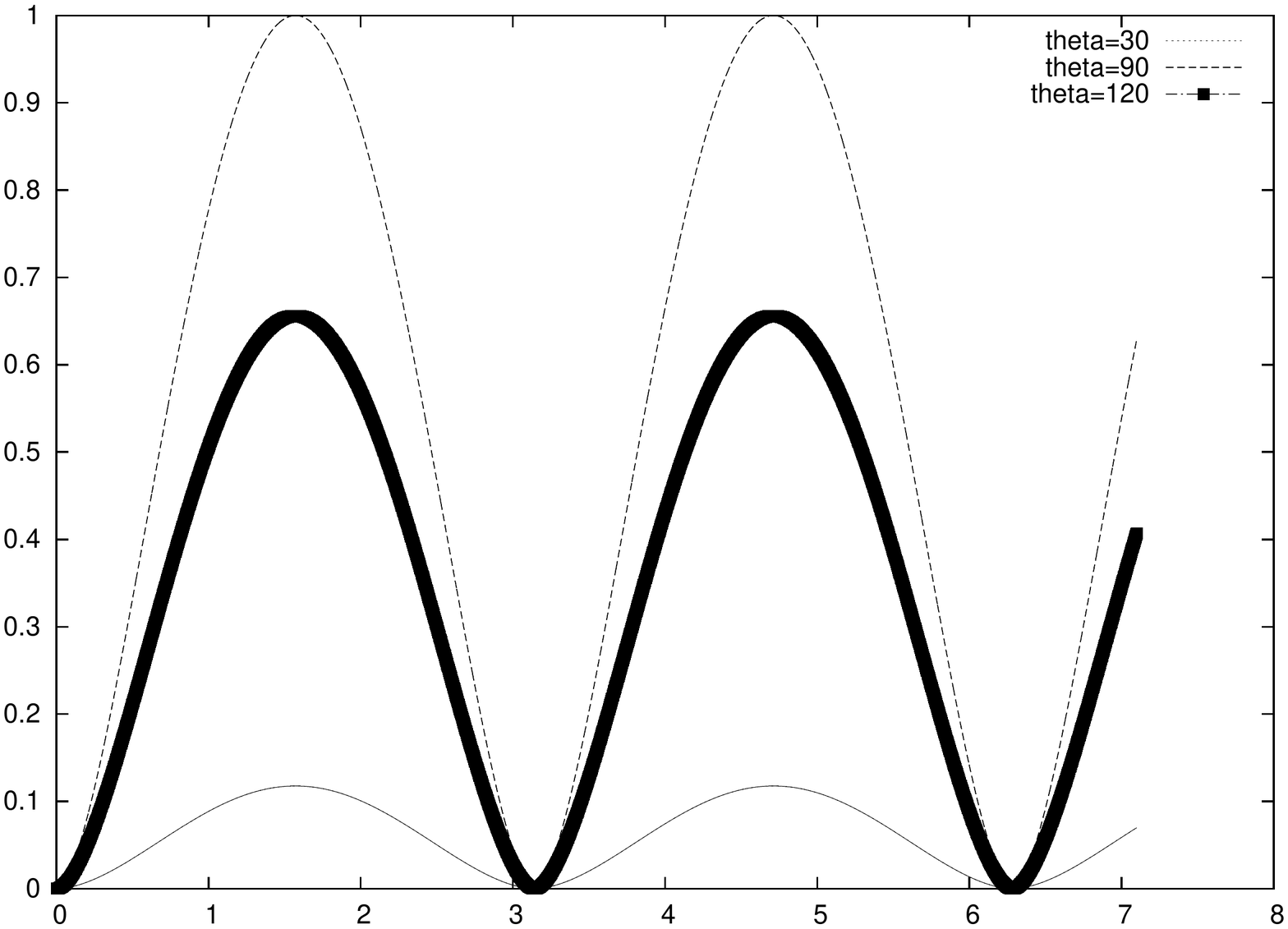}
\caption{[Variation of von Neumann entropy of the partial traces of the system of two two-level atoms with the dimensionless interaction time 
$\Delta_0 t$. We plot the von Neumann entropy and the dimensionless interaction time $\Delta_0 t$ along y-axis and x-axis respectively. The thin line curve (in the bottom) is for $\theta = \pi/6$. The dotted curve (at the top) is for $\theta = \pi/2$ and the thick solid curve 
(at the middle) is for $\theta = 2\pi/3$.]}
\end{center}
\label {fig1}
\end{figure}
Since, in the quantum state of Eq. (\ref{2.3}), both the atoms are on equal footing, the mean spin vectors of both the atoms for the quantum state in Eq. (\ref{2.3}) are equal and has the magnitude (with
$\chi = \tan(\theta/2)$)
\begin{eqnarray}
|\langle\hat{\overrightarrow{J}}_1\rangle| = 
|\langle\hat{\overrightarrow{J}}_2\rangle| =
\frac{1}{2}\sqrt{1 - \sin^2(\Delta_0 t)\sin^4\theta}.
\nonumber\\
\label{2.6}
\end{eqnarray}   

So, using Eqs. (\ref{2.3a3}), (\ref{2.3a4}) and 
(\ref{2.6}), the von Neumann entropy of the partial traces of this system is given as
\begin{eqnarray}
E &=& -\frac{1}{2}\Big{[} 1 + \sqrt{1 - \sin^2(\Delta_0 t)\sin^4\theta} \Big{]}
 log_2 \bigg{[}\frac{1}{2}\Big{(}1 + \sqrt{1 - \sin^2(\Delta_0 t)\sin^4\theta} \Big{)} \bigg{]}
\nonumber\\
&-& \frac{1}{2}\Big{[} 1 - \sqrt{1 - \sin^2(\Delta_0 t)\sin^4\theta} \Big{]}
 log_2 \bigg{[}\frac{1}{2}\Big{(}1 - \sqrt{1 - \sin^2(\Delta_0 t)\sin^4\theta} \Big{)} \bigg{]}.\nonumber\\ 
\label{2.7}
\end{eqnarray}

We show in Fig. 1 the time evolution of this entanglement 
entropy for various values $\theta$. We observe that
the von Neumann entropy of the partial traces increases with $\Delta_0 t$, reaches a maximum and then decreases to zero. This behaviour is repeated with the further increase in $\Delta_0 t$.
This shows that the quantum entanglement shows oscillatroy behaviour with the dimensionless interaction time $\Delta_0 t$ inside the dispersive cavity. The quantum entanglement is maximum when 
$\Delta_0 t = (2n+1)\pi/2$, where n=0, 1, 2, 3.... 
This means that the entanglement entropy is maximum when
\begin{equation}
\frac{g^2\delta}{k^2 + \delta^2} t = (2n +1)
\frac{\pi}{2}.
\end{equation}
Fig. 1 also shows that the entanglement entropy
takes the maximum value 1, when $\theta = \pi/2$.

The angle $\theta$ is related to the probabilities of finding the two atoms in their respective upper and lower states as discussed below.

From the expression of the quantum state given in Eq. (\ref{2.3}) we observe that the probability of finding both the atoms in their respective upper states is 
\begin{equation}
P_1 = \frac{1}{{(1 + |\chi|^2)}^2} = \cos^4\frac{\theta}{2}.
\label{2.9}
\end{equation}
Similarly, the probabilities of finding the first atom in its upper state and the second atom in its lower state and vice-versa is
\begin{equation}
P_2 = P_3 = \frac{|\chi|^2}{{(1 + |\chi|^2)}^2} = \frac{1}{4}
\sin^2\theta.
\label{2.10}
\end{equation}
The probability of finding both the atoms in their lower states is
\begin{equation}
P_4 = \frac{|\chi|^4}{{(1 + |\chi|^2)}^2} = \sin^4
\frac{\theta}{2}.
\end{equation}
These probabilities for $\theta = \pi/6$, 
$\pi/2$, and $2\pi/3$ are given in the following table.
\begin{table}[ht]
\caption{Values of $\theta$, $P_1$, $P_2$, $P_3$ and $P_4$}
\centering
\begin{tabular}{c c c c c}
\hline\hline
$\theta$~~~~~~ &~~~~~~ $P_1$~~~~~~ &~~~~~~ $P_2$~~~~~~ &~~~~~~ $P_3$~~~~~~ & ~~~~~~$P_4$ \\ [0.6ex]
 $\pi/6$ & 0.87 & 0.0625 & 0.0625 & 0.00448\\    
 $\pi/2$ & 0.25 & 0.25 & 0.25 & 0.25\\    
 $2\pi/3$ & 0.0625 & 0.1875  & 0.1875 & 0.5625\\    
 [1 ex]   
\hline
\end{tabular}
\label{table:quant}
\end{table}
\begin{figure}
\begin{center}
\includegraphics[width=10.0cm, angle=0]{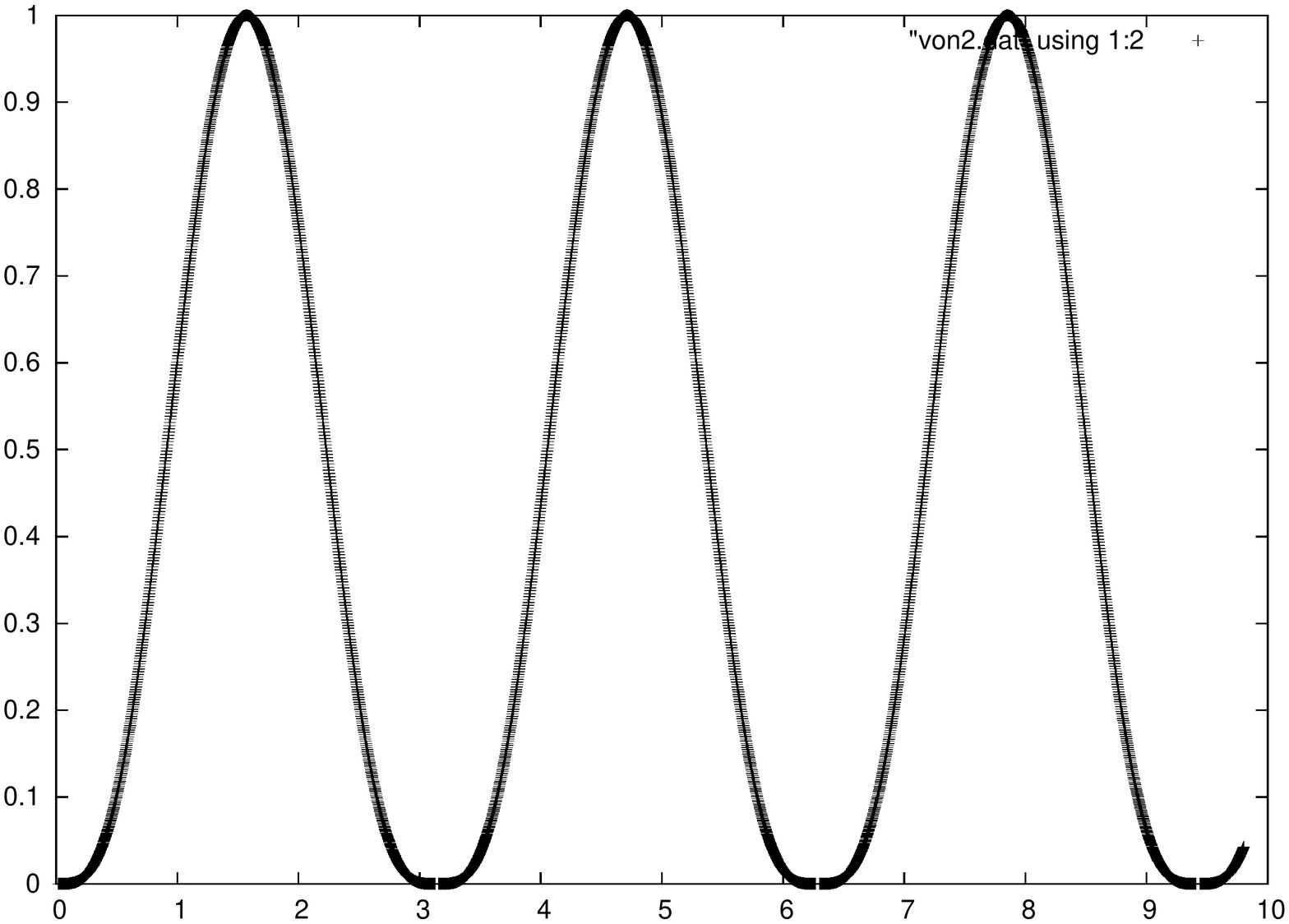}
\caption{[Variation of von Neumann entropy of the partial traces of the system of two two-level atoms with the angle $\theta$. We plot the von Neumann entropy and the angle $\theta$ along y-axis and x-axis respectively. The dimensionless interaction time 
$\Delta_0 t = \pi/2$.]}
\end{center}
\label {fig2}
\end{figure}
We observe from Fig. 1 and Table I, that the von Neumann entropy attains the top most value when
$\theta = \pi/2$ and $P_1 = P_2 = P_3
= P_4 = 0.25$. 

In Fig. 2 we show the variation of the von Neumann entropy of the partial traces with the angle 
$\theta$, keeping fixed the dimensionless interaction time $\Delta_0 t$ at $\pi/2$.
We see that von Neumann entropy increases with increase in $\theta$, goes to a maximum, and then decreases to zero. This behaviour is repeated with further increase in $\theta$. So, quantum entanglement shows oscillatory behaviour with the angle $\theta$.

We, now, proceed to make a contrast of this entanglement entropy with the spin squeezing dynamics of the system.

\section{III. Von Neumann entropy of the partial traces versus spin squeezing}

The spin squeezing of the system of two-level atoms is defined as follows.

If $\hat{J}_x$, $\hat{J}_y$ and $\hat{J}_z$ are the collective spin operators of a system of two-level atoms, then the collective mean spin vector is
\begin{equation}
\langle\hat{\bf J}\rangle = \langle\hat{J}_x\rangle
\hat{i} + \langle\hat{J}_y\rangle
\hat{j} + \langle\hat{J}_z\rangle
\hat{k}.
\label{4.1}
\end{equation}

Now, if $\hat{J}_{1_\perp}$ and $\hat{J}_{2_\perp}$
are the two mutually orthogonal components
of $\hat{\bf J}$ in a plane perpendicular to 
$\langle{\hat{\bf J}}\rangle$, satisfying
\begin{equation}
\Delta J_{1_\perp} \Delta J_{2_\perp} \ge 
\frac{|\langle\hat{\bf J}\rangle|}{2},
\label{4.2}
\end{equation}
 we say that there is spin squeezing in the system if 
$\sqrt{2}\Delta J_{1_\perp}/\sqrt{|\langle\hat{\bf J}\rangle|} < 1$
at the expense of $\Delta J_{2_\perp}$ or,
$\sqrt{2}\Delta J_{2_\perp}/\sqrt{|\langle\hat{\bf J}\rangle|} < 1$
at the expense of $\Delta J_{1_\perp}$.

Now, in general $\langle\hat{\bf J}\rangle$ points
in an arbitrary direction. So, conventionally we rotate our coordinate system $\{x, y, z\}$ to 
$\{x^\prime, y^\prime, z^\prime\}$ such that the mean spin vector $\langle\hat{\bf J}\rangle$ points
along the $z^\prime$ axis. In that case we say that there is spin squeezing in the system if 
\begin{equation}
S_x = \frac{\sqrt{2}\Delta J_{x^\prime}}{\sqrt{|\langle\hat{\bf J}\rangle|}} < 1
\label{4.3}
\end{equation}
 or,
\begin{equation} 
S_y = \frac{\sqrt{2}\Delta J_{y^\prime}}{\sqrt{|\langle\hat{\bf J}\rangle|}} < 1.
\label{4.4}
\end{equation}
\begin{figure}
\begin{center}
\includegraphics[width=10.0cm, angle=0]{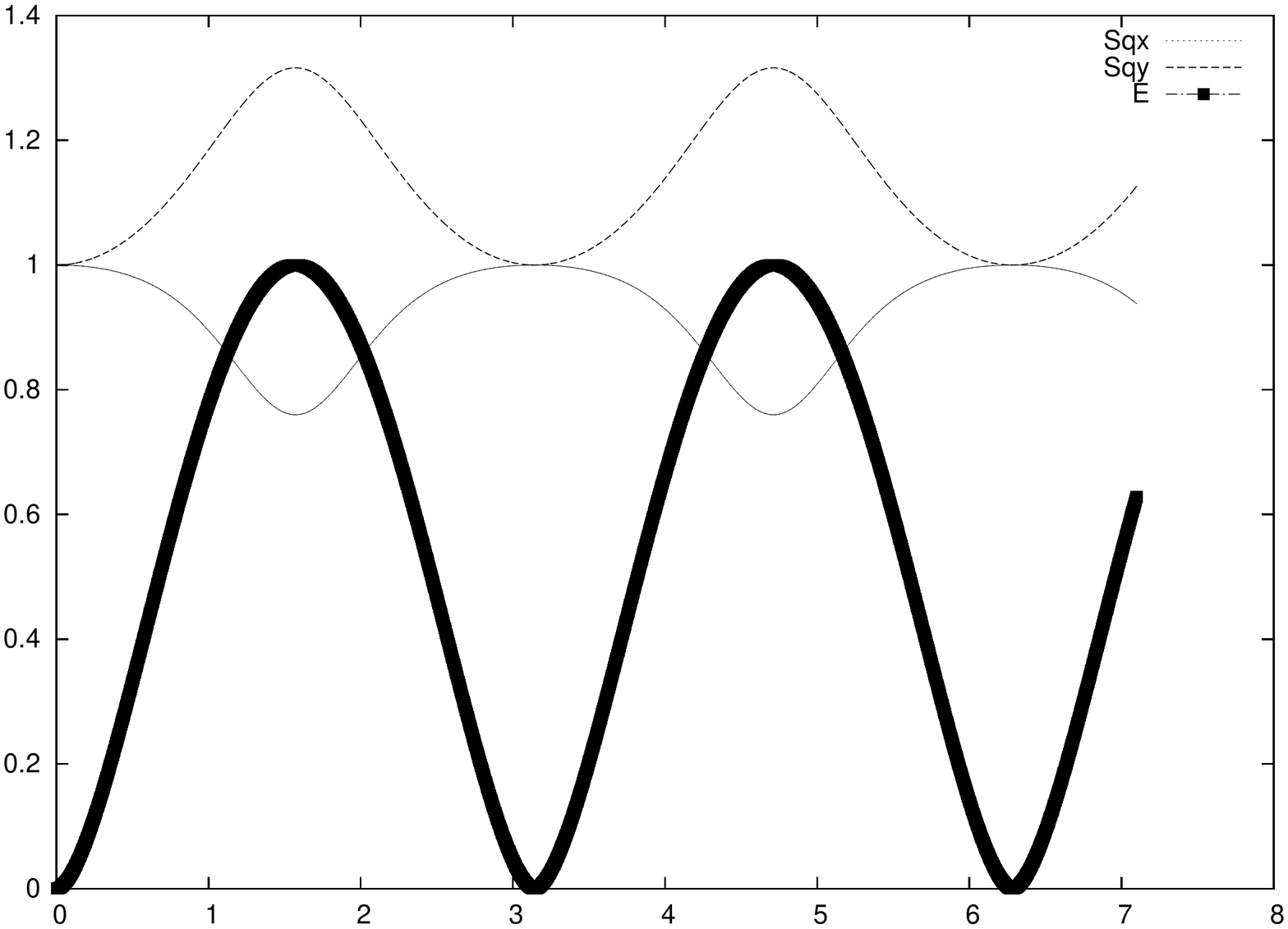}
\caption{[Variation of spin squeezing parameters $S_x$, $S_y$ and entanglement entropy $E$ with  the dimensionless interaction time 
$\Delta_0 t$. We plot $S_x$, $S_y$ and $E$ along 
$y$ axis and $\Delta_0 t$ along $x$ axis. The thick solid curve is E, the thin solid curve is 
$S_x$ and the dotted curve is $S_y$.]}
\end{center}
\label {fig3}
\end{figure}

Spin squeezing properties of this system has been studied in Ref. \cite{Ram}. In this paper we study the dynamics of the von Neumann entropy of this system and make a contrast with the spin squeezing.

Now, in Fig. 3, we plot the spin squeezing parameters $S_x$, $S_y$ and the entanglement entropy $E$ with respect to the dimensionless interaction time $\Delta_0 t$. We observe that when the entanglement entropy goes to maximum, $S_x$ becomes very much less than 1, implying maximum spin squeezing in the $x^\prime$ quadrature. The quantity $S_y$ is never squeezed and it is always greater than 1. $S_y$ goes to maximum when the entanglement entropy $E$ goes to maximum.

We conclude that when the correlations among the two atoms is maximum, showing maximum squeezing, the entanglement entropy is also maximum. Thus, there is a close relationship between spin squeezing and the quantum entanglement.


\begin{thebibliography}{101}
\bibitem{Nielsen}M. A. Nielsen and I. L. Chuang, Quantum Computation and Quantum Information Cambridge University Press, Cambridge, U.K., 2000.  
\bibitem{Raimond}J. M. Raimond, M. Brune, and S. Haroche, Rev. Mod. Phys. {\bf 73}, 565 (2001); 
\bibitem{Solano}E. Solano, G. S. Agarwal, and H. Walther, Phys. Rev. Lett. {\bf 90}, 027903 (2003).
\bibitem{Agarwal} G. S. Agarwal, R. R. Puri, and R. P. Singh, Phys. Rev. A {\bf 56}, 2249 (1997).
\bibitem{Brune} M. Brune, S. Haroche, J. M. 
Raimond, L. Davidovich, and N. Zagury, Phys. Rev. A {\bf 45}, 5193 (1992).
\bibitem{Boissonneault} M. Boissonneault, J. M. Gambetta, and A. Blais, Phys. Rev. A {\bf 77}, 060305 (2008). 
\bibitem{Walther} H. Walther, B. T. H. Varcoe, B. G. Englert, and T. Becker, Rep. Prog. Phys. {\bf 69}, 1325 (2006).
\bibitem{Miller} R. Miller, T. E. Northup, K. M. Birnbaum, A. Boca, A. D. Boozer, and H. J. Kimble, J. Phys. B: At. Mol. Opt. Phys. {\bf 38},
S551 (2005).
\bibitem{Ginzburg} P. Ginzburg, Rev. Phys. {\bf 1}, 120 (2006).
\bibitem{Leek} P. J. Leek, M. Baur, J. M. Fink, R. Bianchetti, L. Steffen, S. Filipp, and A. Wallraff, Phys. Rev. Lett. {\bf 104}, 100504 (2010).
\bibitem{Mekhov} I. B. Mekhov, C. Maschler, and H. Ritsch, Nat. Phys. {\bf 3}, 319 (2007).
\bibitem{Meiser} D. Meiser, and P. Meystre, Phys. Rev. A {\bf 74}, 065801 (2006).
\bibitem{Schuster} D. I. Schuster, A. Fragner, M. I. Dykman, S. A. Lyon, and R. J. Schoelkopf, Phys. Rev. Lett. {\bf 105}, 040503 (2010).
\bibitem{Guo} Y. Guo, R. M. Kroeze, V. D. Vaidya,
J. Keeling, and B. L. Lev, Phys. Rev. Lett. 
{\bf 122}, 193601 (2019).
\bibitem{Kitagawa} M. Kitagawa, M. Ueda, Phys. Rev. A {\bf 47}, 5138 (1993)
\bibitem{Wineland} D. J. Wineland, J. J. Bollinger,
W. M. Itano, Phys. Rev. A {\bf 50}, 67 (1994).
\bibitem{Moya} H. Moya-Cessa, Int. J. Quant. Inf.
{\bf 5}, 149 (2007).
\bibitem{Arevalo} L. M. Arevalo-Aguilar and H. Moya-Cessa, Quant. Sem. Opt. {\bf 10}, 671 (1998).
\bibitem{Moya1} H. Moya-Cessa, S. M. Dutra, J. A. Roversi, A. Vidiella-Barranco, J. Mod. Opt. 
{\bf 46}, 555 (1999).
\bibitem{Radcliffe} J. M. Radcliffe, J. Phys. A {\bf 4}, 313 (1971).
\bibitem{Arecchi} F. T. Arecchi, E. Courtens, R. Gilmore, and H. Thomas, Phys. Rev. A {\bf 6}, 2211 (1972).  
\bibitem{Wotters} W. K. Wootters, Phys. Rev. Lett. {\bf 80}, 2245 (1998). 
\bibitem{Kummer} H. J. Kummer, Int. J. Theor. Phys. {\bf 40}, 1071 (2001).
\bibitem{Chen} J. L. Chen, L. Fu, A. A. Ungar, and X. G. Zhao, Phys. Rev. A {\bf 65}, 044303 (2002). 
\bibitem{Walborn} S. P. Walborn, P. H. Souto Ribeiro, L. Davidovich, F. Mintert, and A. Buchleitner, Phys. Rev. A {\bf 75}, 032338 (2007).
\bibitem{Ram} R. N. Deb, M. S. Abdalla, S. S. Hassan, and N. Nayak, Phys. Rev. A {\bf 73},
053817 (2006).
\end{thebibliography}
\end{document}